\documentclass{webofc}

\usepackage[varg]{txfonts}   
\usepackage{hyperref}
\usepackage{url}
\hypersetup{colorlinks=true,citecolor=blue,urlcolor=blue,linkcolor=blue}
\newcommand{\sqrts}{\sqrt{s_{NN}}}
\begin{document}
\title{Initial conditions and bulk viscosity effects on
$\Lambda$ polarization in high-energy heavy ion
collisions}

\author{\firstname{Andrea} \lastname{Palermo}\inst{1}\fnsep\thanks{\email{andrea.palermo@stonybrook.edu}}
} 

\institute{Center for Nuclear Theory, Department of Physics and Astronomy,
Stony Brook University, Stony Brook, New York 11794-3800, USA 
          }
\abstract{The $\Lambda$ spin polarization is a crucial probe of the gradients of velocity and temperature in the quark-gluon plasma generated in heavy-ion collisions. However, it is still not systematically used to tune hydrodynamic models. In this work, we investigate the influence of different initial conditions and parametrization of the bulk viscosity on $\Lambda$ polarization, showing that they affect the local polarization significantly. These results highlight the impact that the use of local polarization can have on refining theoretical models. Finally, we compare our results, including feed-down corrections, with experimental data from high-energy heavy-ion collisions at STAR and ALICE, and demonstrate the crucial role of bulk viscosity in generating the correct sign of longitudinal polarization at LHC energies.
}
\maketitle
\section{Introduction}
\label{sec:intro}
Polarization of the $\Lambda$ hyperon has become a well-established probe of the quark gluon plasma (QGP) created in high-energy heavy-ion collisions \cite{STAR:2017ckg,STAR:2023nvo,STAR:2018gyt,STAR:2019erd,ALICE:2019onw,ALICE:2021pzu}. Its measurement provides access to finer details of the flow of the QGP compared to standard observables. Indeed, according to the statistical-hydrodynamic model, polarization is sourced by the fluid's vorticity and shear, which are gradients of temperature and four velocity. The role of thermal vorticity as a source of polarization was recognized early on~\cite{Becattini:2013fla,Fang:2016vpj}, whereas the thermal shear has been included in calculations only recently and, critically, it allowed to solve the so-called ``polarization sign puzzle'' \cite{Becattini:2021suc,Liu:2021uhn,Becattini:2021iol,Fu:2021pok}.
We refer the reader to recent reviews on the topic for more details \cite{Becattini:2020ngo,Becattini:2022zvf,Becattini:2024uha}.

The inclusion of the shear term calls for more studies into the sensitivity of polarization to hydrodynamic parameters.
This work is based on ref.~\cite{Palermo:2024tza}, where we have studied the effect of initial conditions, shear and bulk viscosity, and feed-down corrections\footnote{That are corrections to polarization coming from the fact that higher lying resonances decaying to $\Lambda$ particles are themselves polarized.}. We find that the current model provides a good description of available data, and that local polarization is very sensitive to transport coefficients (in particular bulk viscosity) and initial state parameters, and may be used to constrain them.

\section{Polarization in high-energy heavy-ion collisions}
\label{sec:numerics}
\begin{figure}
\centering
\includegraphics[width=0.45\textwidth]{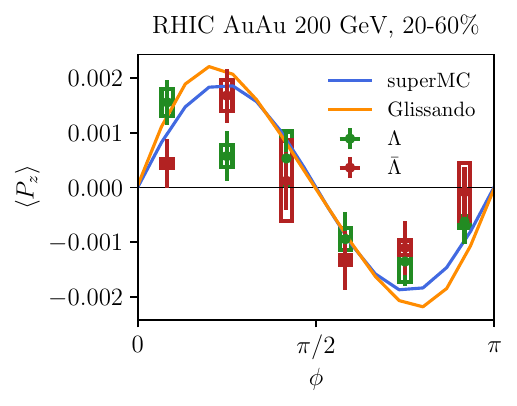}
\includegraphics[width=0.45\textwidth]{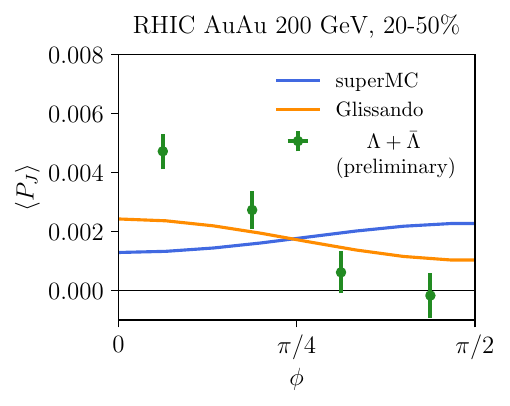}\\
\includegraphics[width=0.45\textwidth]{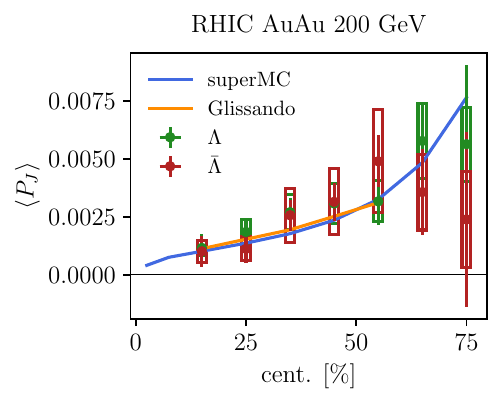}
\caption{Polarization of the $\Lambda$ particle in AuAu collisions at $\sqrts=200$ GeV. Top-left panel: longitudinal polarization (i.e. along the beam direction) as a function of the azimuthal angle. Top-right panel: polarization along the angular momentum direction (transverse polarization) as a function of the azimuthal angle. Low-central panel: transverse polarization as a function of centrality. Data points are taken from refs. \cite{Niida:2018hfw,STAR:2018gyt,STAR:2019erd}}
\label{fig:results}       
\end{figure}

The numerical framework used in this work is based on the codes \texttt{vHLLE} and \texttt{SMASH} \cite{Karpenko:2013wva,Schafer:2021csj,dmytro_oliinychenko_2023_7870822,SMASH:2016zqf}. 
Starting from an initial state, a 3+1D hydrodynamic evolution is carried out by \texttt{vHLLE} \cite{Karpenko:2013wva} until particlization (sometimes referred to as freezeout or hadronization). After that, \texttt{SMASH} \cite{SMASH:2016zqf} samples hadrons from the particlization surface \cite{smash-hadron-sampler} and accounts for subsequent rescatterings and decays. The simulation chain is handled by a dedicated hybrid model \cite{repo}.

 We have used two Initial State (IS) models,
\texttt{superMC} ~\cite{Shen:2020jwv,Alzhrani:2022dpi} and a 3D extension of
\texttt{GLISSANDO} \cite{Rybczynski:2013yba}. Values of free parameters in the IS have been tuned to reproduce hadrons' spectra and elliptic flow. The particlization hypersurface is identified as the constant-energy-density hypersurface $e=0.4$ GeV/fm$^3$. In our simulations, we have used a constant shear viscosity over entropy density $\eta/s$ and a temperature-dependent bulk viscosity over entropy density $\zeta/s$, referred to as ``param III'' in what follows. We refer the reader to \cite{} for further details. 

Throughout this paper, unless otherwise stated, feed-down corrections are always included. We remind the reader that the polarization vector is defined as $P^\mu(p)=S_*^\mu(p)/S$ 
where $S$ is the spin of the particle, and $S_*^\mu(p)$ is its mean spin vector in the particle's rest frame;  for $\Lambda$ hyperons $P^\mu=2 S_*^\mu$. 

We have performed simulations both for $\sqrts = 200$ GeV AuAu collisions, and $\sqrts=5020$ GeV PbPb collisions, corresponding to the STAR and ALICE experiments. However, due to space constraints, we show results only for the former in Figure \ref{fig:results}.
Our results describe well the longitudinal component of polarization $P_z$ (the projection along the beam direction), confirming previous findings \cite{Becattini:2021iol}. However, an interesting difference between our two initial states becomes apparent in the transverse polarization (the projection along the global angular momentum of the QGP). Despite yielding the same longitudinal polarization, the transverse polarization $P_J$ as a function of the azimuthal angle differs significantly, and the data seem to favour \texttt{GLISSANDO}. Such a difference, however, disappears when comparing the models as a function of centrality, shown in the lower central panel of Fig.~\ref {fig:results}. This observation shows that the azimuthal dependence of $P_J$ can strongly constrain initial state models. Similar behavior is observed in PbPb collisions, indicating that these findings are likely robust across different collision energies.

\section{Effect of feed down, viscosity, and initial conditions}
\label{sec:visc}
The main results of this investigation are the effects of feed down, transport coefficients, and initial state on the $\Lambda$ spin polarization. 

Starting with the feed-down corrections, we have studied the polarization of mother resonances, namely $\Sigma^*$ and $\Sigma_0$, which decay into $\Lambda$ through strong and electromagnetic interactions. The momentum-dependent polarization inherited by the $\Lambda$ in the decay has been computed according to ref.~\cite{Becattini:2019ntv}, see also \cite{Becattini:2016gvu,Xia:2019fjf} for related studies. The total polarization has been computed as the average of the primary and the feed down $\Lambda$s, each channel being weighted by the respective multiplicity. We find that the feed down produces a small effect on spin polarization, being a $3\div 10\%$ effect at most. This is in agreement with previous literature, where only thermal vorticity was used as a source of polarization.\cite{Xia:2019fjf,Becattini:2019ntv}.
\begin{figure}
    \centering
\includegraphics[width=0.45\textwidth]{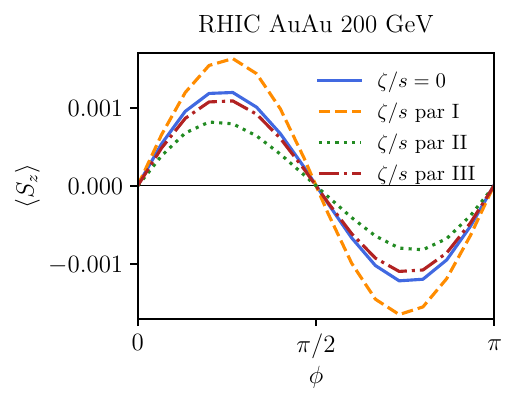}
\includegraphics[width=0.45\textwidth]{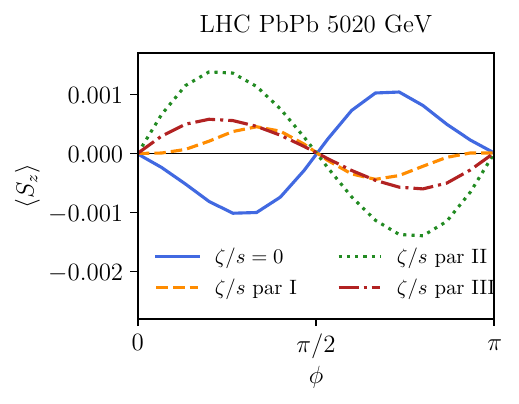}
    \caption{Longitudinal component of the spin vector as a function of 
    the azimuthal angle $\phi$ at in AuAu $\sqrts = 200$ collisions (left panel) and PbPb $\sqrts = 5020$ collisions (right panel) for various bulk viscosity parametrizations.}
    \label{fig: bulk params}
\end{figure}

Shear viscosity and bulk viscosity effects have also been studied. We have used different values of a constant shear viscosity over entropy ratio, $\eta/s \in \{0,0.08,0.16\}$, with very small modification to the longitudinal polarization and no significant change to the transverse polarization $P_J$. This is in line with the findings of other authors \cite{}. In contrast, we find that bulk viscosity is critical to reproduce the correct sign of longitudinal polarization, especially at the energies probed by ALICE, $\sqrts=5020$ GeV ~\cite{Palermo:2022lvh}. We have tested three different parametrizations, dubbed ``Parametrization I'' ~\cite{Ryu:2017qzn}, ``Parametrization II'' ~\cite{Schenke:2019ruo}, and ``Parametrization III'' \cite{Schenke:2020mbo}, comparing them to the case of a vanishing bulk. We refer the reader to the original sources and to our main publication for further details.
Figure \ref{fig: bulk params} shows our results, the main finding being that the presence of bulk viscosity changes the sign of local longitudinal polarization in PbPb collisions at $5020$ GeV, whereas the effect at lower energies, despite being still more significant than the one of shear viscosity, remains rather contained. It's hard to single out the reason for this effect. An analysis of independent velocity-gradients contributions to polarization has been performed in \cite{}, with the conclusion that angular velocity and acceleration receive larger modifications from bulk viscosity at LHC energies rather than at RHIC.  
Possibly, the higher temperatures and longer lifetime of the QGP make the role of bulk viscosity more important at LHC. 

\begin{figure}
    \centering
\includegraphics[width=0.45\textwidth]{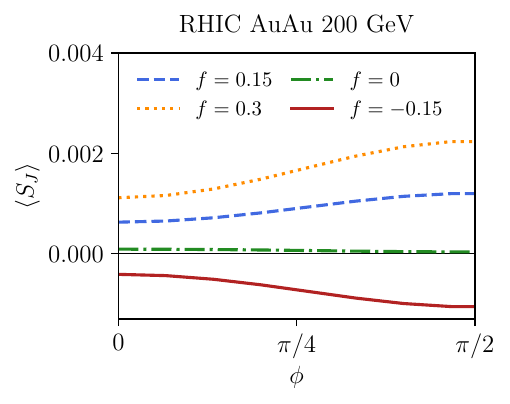}
\includegraphics[width=0.45\textwidth]{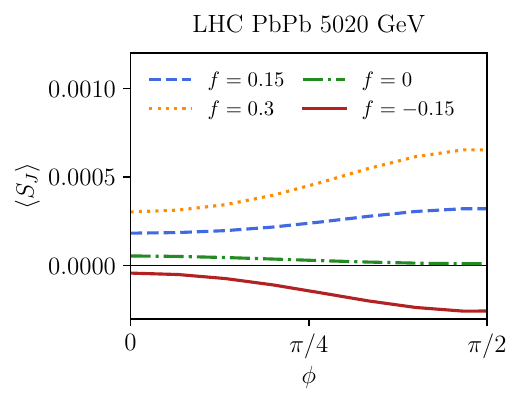}
    \caption{$S_J$ component of the $\Lambda$ particle's mean spin for different values \texttt{superMC}'s $f$ free parameter, at RHIC and LHC energies in the left and right panels respectively.} 
    \label{fig:fs}
\end{figure}
Finally, we address the sensitivity of polarization to the initialization of the energy-momentum tensor. This turns out to explain the difference between \texttt{GLISSANDO} and \texttt{superMC} concerning the $P_J(\phi)$ component of polarization. In \texttt{superMC}, a free parameter $f$ determines the initialization of the components of the energy momentum tensor as $T^{\tau\tau}=\rho \cosh(f\,y_{CM})$  and $T^{\tau\eta}=\frac{\rho}{\tau}\sinh(f\,y_{CM})$, where $\rho$ is the local energy density distribution and $y_{CM}$ is the center of mass rapidity, whereas in \texttt{GLISSANDO} only the $\tau\tau$-component is initialized, setting effectively $f$ to zero. Figure \ref{fig:fs}, shows how different values of the parameter $f$, affect the transverse polarization. For $f=0$ we obtain a decreasing polarization with the azimuthal angle, as in the \texttt{GLISSANDO} model shown in figure \ref{fig:results}\footnote{The quantitative difference between the two models with $f=0$ lies in the fact that other free initial parameters have not been modified.}. This shows that the local transverse polarization can be an important probe for the initial state. 

\section{Conclusions}
\label{sec:conclusions}

In conclusion, we have analysed $\Lambda$ spin polarization in Au-Au collisions at $\sqrts=200$ GeV and in Pb-Pb collisions at $\sqrts=5020$ GeV. We have shown that transverse polarization is sensitive to the initial longitudinal flow, while longitudinal polarization at LHC energies is extremely sensitive to bulk viscosity.
These results demonstrate that spin polarization can be an effective probe of both the initial conditions and the transport coefficients 
of the Quark-Gluon plasma.

 \bibliography{biblio} 

\begin{thebibliography}{33}
\bibitem{repo}
SMASH-vHLLE hybrid with Python, the repository will be made available upon
  publication.
\end{thebibliography}

\end{document}